\documentclass{article}
\usepackage[twocolumn]{emulateapj}
\usepackage{epsf}

\def\plotone#1{\centering \leavevmode
\epsfxsize= 0.8\columnwidth \epsfbox{#1}}
\def\plotonecita#1{{}}
\def\plotonecfpa#1{{\plotone{#1}}}
\def\spose#1{\hbox to 0pt{#1\hss}}
\def\lta{\mathrel{\spose{\lower 3pt\hbox{$\mathchar"218$}}
     \raise 2.0pt\hbox{$\mathchar"13C$}}}
\def\gta{\mathrel{\spose{\lower 3pt\hbox{$\mathchar"218$}}
     \raise 2.0pt\hbox{$\mathchar"13E$}}}
\def\be{\begin{equation}}
\def\ee{\end{equation}}
\def\bea{\begin{eqnarray}}
\def\eea{\end{eqnarray}}

\begin{document}
\title{Computing Challenges of the Cosmic Microwave Background}
\author{J. Richard Bond\altaffilmark{1}, Robert G. Crittenden\altaffilmark{2}, 
Andrew H. Jaffe\altaffilmark{3} and Lloyd Knox\altaffilmark{4}}
\affil{1,2. Canadian Institute for Theoretical Astrophysics, Toronto, ON M5S
  3H8, CANADA \\
3, Center for Particle Astrophysics,
  301 LeConte Hall, University of California, Berkeley, CA 94720\\
4. Astronomy \&
  Astrophysics Center, 5640 S.\ Ellis Ave., Chicago, IL 60637}

\altaffiltext{1}{E-mail: bond@cita.utoronto.ca}
\altaffiltext{2}{E-mail: crittend@cita.utoronto.ca}
\altaffiltext{3}{E-mail: jaffe@cfpa.berkeley.edu}
\altaffiltext{4}{E-mail: knox@flight.uchicago.edu}

\begin{abstract}
The Cosmic Microwave Background (CMB) encodes information on the
origin and evolution of the universe, buried in a fractional
anisotropy of one part in $10^{5}$ on angular scales from arcminutes
to tens of degrees.  We await the coming onslaught of data from
experiments measuring the microwave sky from the ground, from balloons
and from space. However, we are faced with the harsh reality that
current algorithms for extracting cosmological information cannot
handle data sets of the size and complexity expected even in the next
few years.  Here we review the challenges involved in understanding
this data: making maps from time-ordered data, removing
the foreground contaminants, and finally estimating the power spectrum
and cosmological parameters from the CMB map.  If handled naively, the
global nature of the analysis problem renders these tasks effectively
impossible given the volume of the data.  We discuss possible
techniques for overcoming these issues and outline the many other
challenges that wait to be addressed.

[Invited article for {\em Computing in Science and Engineering.}]

\vskip 0.2in
\end{abstract}


\section{The CMB}

\subsection{Historical Overview}
 
The detection of the cosmic microwave background (CMB) in 1965 stands
as one of the most important scientific discoveries of the century,
the strongest evidence we have of the Hot Big Bang model. We know from
the COBE satellite that it is an almost perfect blackbody with
temperature $2.728\pm 0.004$ K, with expected tiny spectral
distortions only very recently discovered.  Once the CMB was
discovered, the search was on for the inevitable angular fluctuations
in the temperature, which theorists knew would encode invaluable
information about the state of the universe at the epoch when big bang
photons decoupled from the matter. This occurred as the universe cooled
sufficiently for the ionized plasma to combine into hydrogen and
helium atoms. This epoch was a few hundred thousand years after the
Big Bang, at a redshift $z\sim 1000$ when the universe was a factor of
a thousand smaller than it is today.

Theorists led the experimenters on a merry chase, originally predicting
the fractional temperature fluctuation level would be $10^{-2}$, then in
the seventies $10^{-3}$, then $10^{-5}$, where it has been since the
early eighties, when the effects of the dark matter which dominates the
mass of the universe were folded into the predictions. Fortunately the
experimenters were persistent, and upper limits on the anisotropy
dropped throughout the eighties, leaving in their wake many failed ideas
about how structure may have formed in the universe.

A major puzzle of the hot big bang model was how regions that would
not have been in causal contact at redshift $z \sim 1000$ could have the
same temperature to such a high precision. This led to the theory of
inflation, accelerated expansion 
driven by the energy density of a
scalar field, dubbed the inflaton, 
in which all of the universe we can
see was in contact a mere $\lta 10^{-33}$ seconds after the big
bang. It explained the remarkable isotropy of the CMB 
and had a natural byproduct: quantum 
oscillations in the scalar field could have generated the density
fluctuations that grew via gravitational instability to create the
large scale structure we see in the universe around us. This theory,
plus the hypothesis that the dark matter was made up of elementary
particle remnants of the big bang, led to firm predictions of the
anisotropy amplitude. In the eighties, competing theories 
arose, one of which still survives: that topologically
stable configurations (defects, such as cosmic strings) of exotic particle fields 
arising in phase transitions
could have formed in the early universe and acted as seeds for the
density fluctuations in ordinary matter.

Immediately following the headline-generating detection of
anisotropies by COBE [\cite{bennett}] in 1992 at the predicted $\sim
10^{-5}$ level, many ground and balloon
experiment began seeing anisotropies over a broad range of angular
scales. The emerging picture from this data has sharpened our
theoretical focus to a small group of surviving theories,
such as the inflation idea.  The figures in this article tell the
story of where we go from here.  Fig.~\ref{fig:allsky} shows a
realization of how the temperature fluctuations would look on the sky
in an inflation-based model, at the $\sim 7^\circ$ resolution of the
COBE satellite and what would be revealed at essentially full
resolution.  One sees not only the long wavelength ups and downs that
COBE saw, but also the tremendous structure at smaller scales in the
map.

One measure of this is the {\it power spectrum} of the temperature
fluctuations, denoted by $C_\ell$, a function of angular wavenumber
$\ell$, or, more precisely, the multipole
number in a spherical harmonic expansion. Fig.~\ref{fig:CLtheory}
shows typical predictions of this for the inflation and defect
theories, and contrasts it with the best estimate from all of the
current data. The ups and downs in $\ell$-space are associated with
sound waves at the epoch of photon decoupling. The damping evident at
high $\ell$ is a natural consequence of the viscosity of the gas as
the CMB photons are released from it. The flat part at low $\ell$ is
associated with ripples in the past light cone arising from
gravitational potential fluctuations that accompany mass
concentrations. All of these effects are sensitive to cosmological
parameters, {\it e.g.}, the densities of baryons and dark matter, the
value of the cosmological constant, the average curvature of the
universe, and parameters characterizing the inflation-generated
fluctuations. If the spectrum can be measured accurately enough
experimentally, such cosmological parameters can also be determined
with high accuracy.  For a review of CMB science see [\cite{btw}].

\begin{figure}
\begin{center}
\plotonecfpa{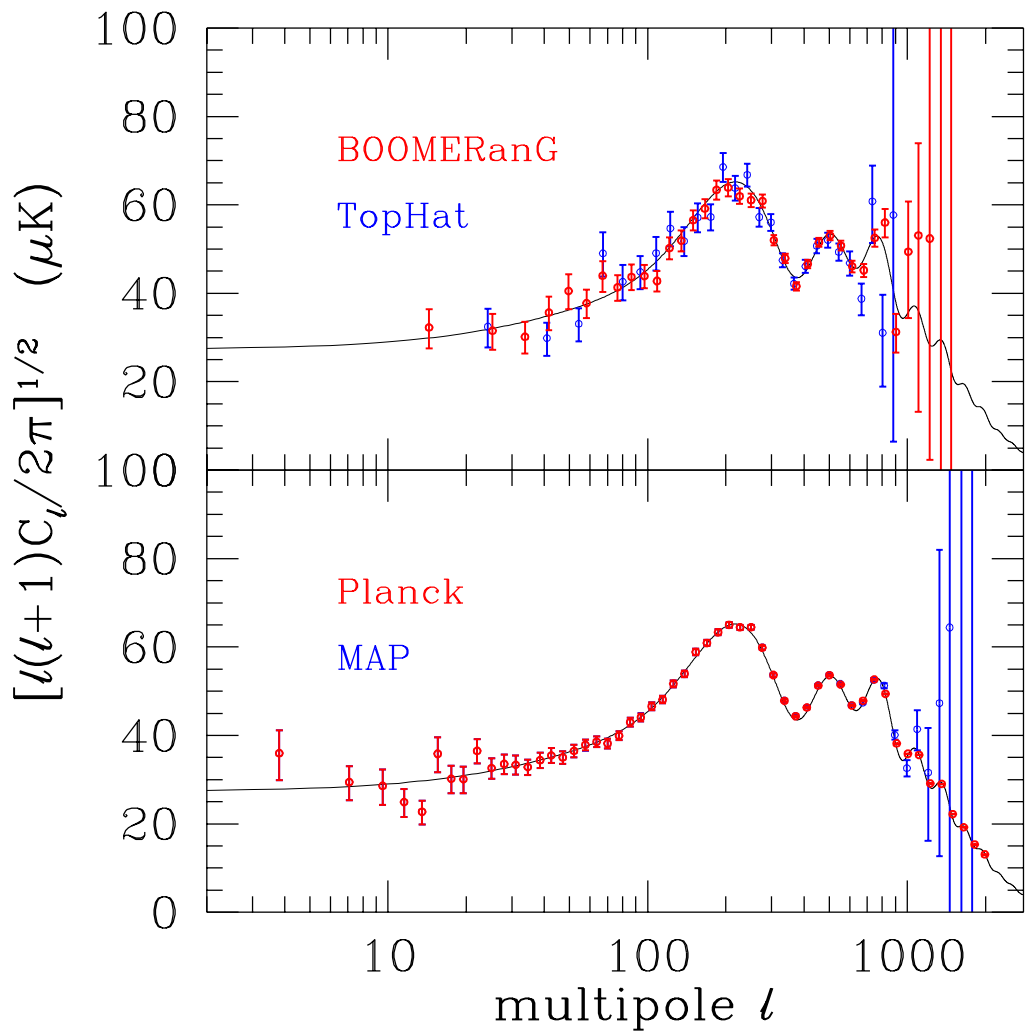}
\caption[pointing]{\baselineskip=10pt\small 
The plot shows simulations
of power spectra and their error bars for four upcoming experiments,
with two long duration balloon missions at top, and the two satellite
missions in the lower figure.  The techniques of \S~\ref{sec:paramest}
were used, but foregrounds were ignored and the noise was assumed to
be homogeneous. While the larger scale measurements at low $\ell$ will
not improve much beyond what is already known from COBE, the smaller
scale measurements will improve dramatically with time.  Note that
where the MAP error bars do not appear, they are identical to
Planck's.}
\label{fig:clproj}
\end{center}
\end{figure}

Once it became clear that there was something to measure, the race was
on to design high-precision experiments that would cover large areas
of the sky at the fine resolution needed to reveal all this structure
and the wealth of information it encodes. These include ground-based
interferometers and long duration balloon (LDB) experiments (flying for 10 days
vs. 10 hours for conventional balloon flights), as well as the use of
large arrays of detectors.  NASA will launch the Microwave Anisotropy
Probe (MAP) [\cite{MAP}] satellite in 2000 and ESA will launch the
Planck Surveyor [\cite{Planck}] around 2006. They will each spend a
year or two mapping the full sky.  Fig.~\ref{fig:clproj} gives an idea
of how well we think that the LDB and satellite experiments can do in
determining $C_\ell$ if everything goes right. Theorists have also
estimated how well the cosmological parameters that define the
functional dependence of $C_\ell$ in inflation models can in principle
be determined with these experiments. In one exercise that allowed a
mix of nine cosmological parameters to characterize the space of
inflation-based theories, COBE was shown to determine one combination
of them to better than 10\% accuracy, LDBs and MAP could determine six,
and Planck seven. MAP would also get three combinations to 1\% accuracy, and
Planck seven!  This is the promise of a high-precision cosmology as we
move into the next millennium.

\subsection{Experimental Concerns}
 
CMB anisotropy experiments often involve a number of microwave and
sub-millimeter detectors covering at least a few frequencies, located
at the focal plane of a telescope. The raw data comes to us as noisy
time-ordered recordings of the temperature for each frequency channel,
which we shall refer to as timestreams, along with the pointing vector
of each detector on the sky. The resolution of the experiment is
usually fixed by the size of the telescope and the frequency of the
radiation one looks at.  We must learn from the data itself almost
everything about the noise and the many signals expected, both wanted
and unwanted, with only some guidance from other astrophysical
observations. We shall see that to a large degree this appears to be a
well-posed problem in Bayesian statistical analysis.

The major data products from the COBE anisotropy experiment were six
maps, each with 6144 pixels, derived from six timestreams, one for
each detector. The timestream noise was Gaussian, which
translated into correlated Gaussian noise in the maps. Much effort
went into full statistical analyses of the underlying sky signals,
most often under the hypothesis that the sky signal was a Gaussian
process as well. The amount of COBE data was at the edge of what could
be done with 1992 workstations. The other experiments used in the
estimate of the power spectrum in Fig.~\ref{fig:CLtheory} had less
data, and full analysis was also feasible. We are now entering a new
era: LDB experiments will have up to two orders of magnitude more
data, MAP three and Planck four. For the forecasts of impressively
small $C_\ell$ errors to become reality, we must learn to deal with
this huge volume of data.
In this article, we discuss the computational challenges
associated with current methods for going from the timestreams to
multi-frequency sky maps, and for separating out from these maps
of the different sky signals. 
Finally, from the CMB map and its statistical properties,
cosmological parameters can be derived. 
To illustrate the techniques, we use them to find estimates of $C_\ell$. 
This represents an
extreme form of data compression, but from which cosmological
parameters and their errors can finally be derived.

As we shall discuss at considerable length in this article, the analysis
procedure we will describe is necessarily {\em global}; that is, making
the map requires operating on the entire time-ordered data, and estimating the
power spectrum requires analyzing the entire map at once.  This is due
to the statistically-correlated nature of both the instrumental noise
and the expected CMB sky signal which links up measurements made at one
point with those made at all others.

\section{What signals do we expect?} 

Of the signals we know are present, there are of course the {\it
primary} CMB fluctuations from the epoch of photon decoupling that we have
already discussed, the primary goal of this huge worldwide
effort. There are also {\it secondary} fluctuations of great interest
to cosmologists arising from nonlinear processes at lower redshift:
some come from the epoch of galaxy formation and 
some from scattering of CMB photons by hot gas in clusters of
galaxies. Extragalactic radio sources are another nontrivial
signal. On top of this, there are various 
emissions from dust and gas in our Milky Way galaxy.  While these  
are foreground nuisances to cosmologists, they are signals of passionate
interest to interstellar medium astronomers.  Fortunately these
signals have very different dependences on frequency
(Fig.~\ref{fig:frequencyforegrounds}), and, as we now know, rather
statistically distinct sky patterns
(Fig.~\ref{fig:spatialforegrounds}).

\begin{figure}  
\begin{center}
\plotone{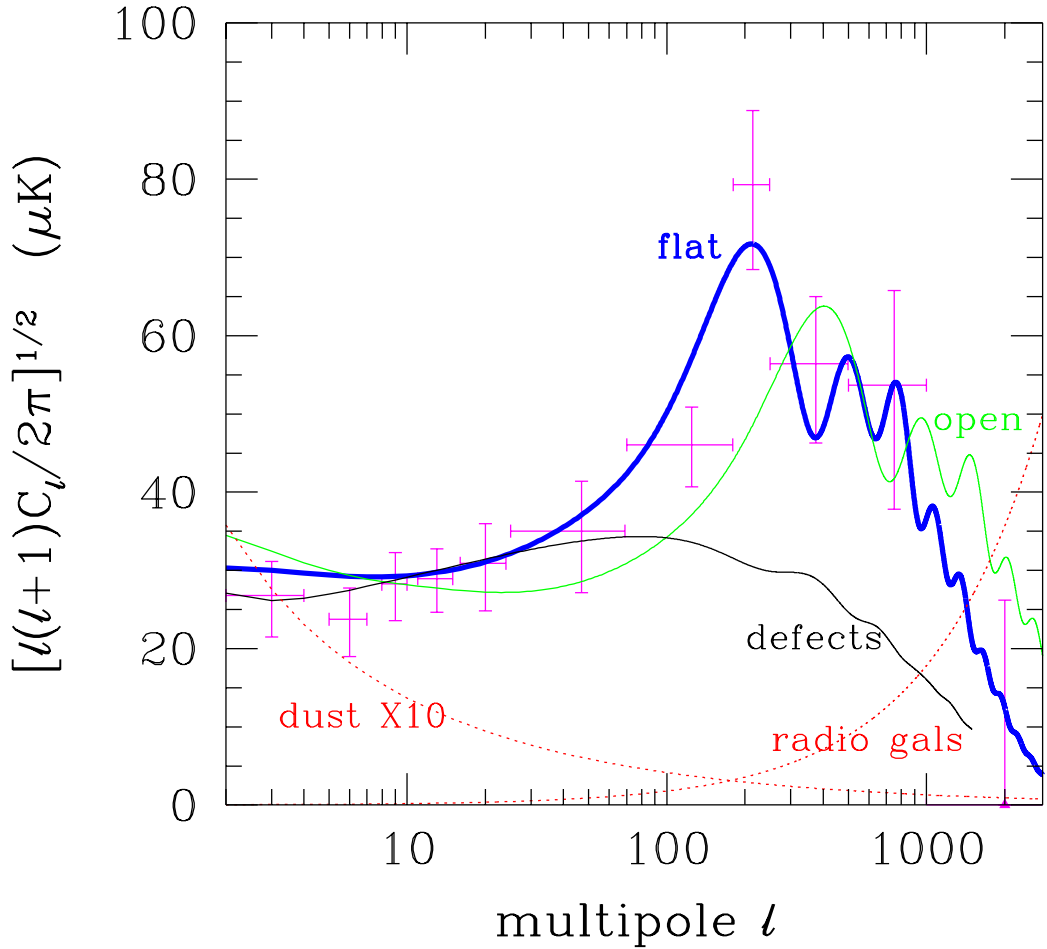 }
\caption[CLtheory]{\baselineskip=10pt\small Theoretical power spectra
as a function of multipole number $\ell$ for a few theories (solid
lines) are contrasted with the power in bands derived from the current
data.  Large angles are at low multipole number, small at high. The
multiplier of $C_\ell$ is chosen to show power per logarithmic $\ell$ 
bin, so equal power on all scales would appear as a
horizontal line.  The flat model shown is the current best fit to the
cosmological data, involving a mix of cold dark matter, baryons and a
sizable cosmological constant. The height of the $C_\ell$ (acoustic)
peaks is sensitive to changes in cosmological parameters; {\it e.g.},
increasing the baryon content raises them. The open cosmological model
has a similar amount of cold dark matter and baryons. The shift of the
peaks to higher $\ell$ is a consequence of the negatively curved
geometry.  Also shown is a sample $C_\ell$ for the defect model in
which structure forms as a response to topological field
configurations in early universe phase transitions. While still
tentative, the data favor the flat model over the open and defect
models, and future measurements will be able to distinguish theories
that differ by less than the thickness of the blue line.  Foregrounds,
shown in red, typically have very different angular spectra: either
dominated by large structures like the galactic dust maps, or by point sources
like the radio galaxies.}
\label{fig:CLtheory}
\end{center}
\end{figure}

We know how to calculate in exquisite detail the statistics of the 
primary signal for
the various models of cosmic structure formation. The fluctuations are
so small at the epoch of photon decoupling that linear perturbation
theory is a superb approximation to the exact non-linear evolution
equations.  The simplest versions of the
inflation theory predict that the fluctuations from the quantum noise
form a Gaussian random field.  Linearity implies that this
translates into anisotropy patterns that are drawn from a Gaussian
random process and which can be characterized solely by their power spectrum.
Thus our emphasis is on confronting the theory with the
data in the power spectrum space, as in
Fig.~\ref{fig:CLtheory}. Primary anisotropies in defect theories are
more complicated to calculate, because non-Gaussian patterns are
created in the phase transitions which evolve in complex ways and for
which large scale simulations are required, a computing challenge we
shall not discuss in this article. In both theories, algorithmic
advances have been very important for speeding-up the computations of
$C_\ell$.

The {\it secondary} fluctuations involve nonlinear processes, and the
full panoply of $N$-body and gas-dynamical cosmological simulation
techniques discussed in this volume are being brought to bear on the
calculation. Non-gaussian aspects of the predicted patterns are
fundamental, and much beyond $C_\ell$ is required to specify
them. Further, some secondary signals, such as radiation from dusty
star-burst regions in galaxies, are too difficult to calculate from
first principles, and statistical models of their distribution must be
guided by observations. At least for most CMB experiments, they can be
treated as point sources, much smaller than the observational
resolution. The {\it foreground} signals from the interstellar medium
are also non-Gaussian and not calculable.  They must be modeled from
the observations and have the added complication of being extended
sources.

For each signal $T$ present, there is therefore a theoretical ``prior
probability'' function specifying its statistical distribution, ${\cal
  P}(T\vert {\rm th} )$.  A Gaussian ${\cal P}(T\vert {\rm th} )$ has
the important property that it is completely specified by the two-point
correlation function which is the expectation value of the product of
the temperature in two directions ${\bf \hat q}$ and ${\bf \hat q'}$
on the sky, $\langle T({\bf \hat q}) T({\bf \hat q'})\rangle$. For
non-Gaussian processes an infinite number of higher order temperature
correlation functions are needed in principle. The inflation-generated or
defect-generated temperature anisotropies are also usually statistically
{\it isotropic}, that is, the $N$-point correlation functions are
invariant under a uniform rotation of the $N$ sky vectors ${\bf \hat
  q}$. This implies $\langle T({\bf \hat q}) T({\bf \hat q'})\rangle$ is
a function only of the angular separation.  If the
temperature field is expanded in spherical harmonics $Y_{\ell m}({\bf
  \hat{q}})$, then the two-point function of the coefficients $a_{\ell
  m}$ is related to $C_\ell$ by \be \langle a_{\ell m} a_{\ell' m'}^*
\rangle = C_\ell \delta_{\ell\ell'}\delta_{mm'}, \quad {\rm where}\  T({\bf
  \hat{q}}) = \sum_{\ell m} a_{\ell m} Y_{\ell m}({\bf \hat{q}}),
\label{eqn:sph-trn}
\ee 
so the correlation function is related to the $C_\ell$ by
\begin{equation}
  \label{eqn:TT}
\langle T({\bf \hat q}) T({\bf \hat q'})\rangle =
\sum_\ell {2\ell+1\over4\pi}
C_\ell P_\ell({\bf\hat q}\cdot{\bf\hat q}),
\end{equation}
where $P_\ell(x)$ is a Legendre polynominal.
Just as a Fourier wavenumber $k$ corresponds to a scale
$\lambda\sim2\pi/k$, the spherical-harmonic coefficients correspond to
an angular scale $\theta\sim180^\circ/\ell$. Figure~\ref{fig:CLtheory}
shows $C_\ell$ for two different cosmologies given the same primordial
theory; we plot $\ell(\ell+1)C_\ell/(2\pi)$ since at high $\ell$ it
gives the power per logarithmic bin of $\ell$.

A nice way to think about Gaussian fluctuations is that for a given
power spectrum, they distribute this power with the smallest
dispersion.  Temperature fluctuations are typically within $\pm 2
\sigma$ and rarely exceed $3\sigma$, where $\sigma$ is {\it rms}
amplitude. Such is the map in Fig.~\ref{fig:allsky}.  Since the term 
non-Gaussian
covers all other possibilities, it may seem impossible to characterize,
but the way the greater dispersion often manifests itself is that the
power is more concentrated, {\it e.g.} in extended hot and/or cold spots
for the galactic foregrounds, and point-like concentrations 
for the extragalactic
sources, as is evident in Fig.~\ref{fig:spatialforegrounds}.

Although we may marvel at how well the basic inflation prediction from
the 1980's is doing relative to the current data in
Fig.~\ref{fig:CLtheory}, it will be astounding is if no anomalies
are found in the passage from those large error bars to the much
smaller ones of Fig.~\ref{fig:clproj} and human musings about such
exotic ultra-early universe processes are confirmed.

\section{What is coming?} 

The new CMB anisotropy data sets will come from a variety of platforms:
large arrays of detectors on the ground or on balloons, long duration
balloons (LDBs), ground-based interferometers and satellites.  Most of
these experiments measure the sky at anywhere between 3 to 10 photon
frequencies, with several detectors at each frequency.  With detector
sampling rates of about 100 Hz and durations of weeks to years, the
raw data sets range in size from Gigabytes to nearly Terabytes.

Another measure of the size of a data set is the number of resolution
elements, or beam-size pixels, in the maps that are derived from the
raw data.  Over the next two years, LDBs and interferometers will
measure between $10^4$ to $10^5$ resolution elements, which is an
impressive improvement upon COBE/DMR's $10^3$ elements.  NASA's MAP
satellite will measure the whole sky with $12'$ resolution in its
highest frequency channel, resulting in CMB maps with $\sim 10^6$
resolution elements.  The Planck Surveyor has $5'$ resolution, that of the lower panel of Fig.~\ref{fig:allsky}, and will create
maps with $\sim 10^7$ resolution elements.

In Fig.~\ref{fig:clproj}, forecasts of power spectra and their errors
for TopHat and BOOMERanG (two LDB missions) and MAP and Planck are
given.  These results ignore foregrounds and assume maps have
homogeneous noise, and thus are highly idealized. Extracting the
angular power spectrum from such large maps presents a formidable
computing challenge.  Except for the complication of being on a
sphere, the difficulties are those shared with the more usual problem
of power spectrum estimation in flat spaces; in general, it is an
$O(m_p^3)$ process, where $m_p$ is the number of pixels in the map.
What makes the process $O(m_p^3)$ is either matrix inversion or
determinant evaluation, depending on the particular implementation.
(In special cases, the Fast Fourier Transform is a particularly
elegant matrix factorization, reducing the operations count from
$O(m_p^3)$ to $O(m_p \ln m_p)$, but it is not generally applicable.)
In addition to the operations count, storage is also a challenge,
since the operations are manipulations of $m_p \times m_p$ matrices.
For example, the noise correlation matrix for a megapixel map requires
2000 Gbytes for single precision (four byte) storage!

\section{Following the thread of the data:} 

Conceptually, the process of extracting cosmological information from a
CMB anisotropy experiment is straightforward. 
First, maps of microwave emission at the observed wavelengths are
extracted from the lengthy time-ordered data; these are the
maximum-likelihood estimates of the sky signal given a noise
model. Then, the various physical components are separated:
solar-system contamination, galactic and extragalactic foregrounds,
and the CMB itself.
Finally, given the CMB map, we can find the maximum-likelihood
power spectrum, $C_\ell$, from which 
the underlying cosmological
parameters can be computed. 
This entire data
analysis pipeline can be unified in a Bayesian likelihood formalism.
Of course, this pipeline is complicated by
the correlated nature of the instrumental noise,  
by unavoidable systematic effects and 
by the non-Gaussian nature of the various sky signals.

\subsection{From the instrument to the maps\ldots} 

Experiments measure the microwave emission from the sky convolved with
their {\em beam}.  Measurements of different parts of the sky are
often combined using complicated difference schemes, called {\em
chopping patterns}.  For example, while the Planck Surveyor will
measure the temperature of a single point on the sky at any given time, 
MAP and COBE measure the temperature difference
between two points.  The purpose of
these chops is to reduce the noise contamination between samples,
which can be large and may have long-term drifts and other
complications.  Observations are repeated many times over the
experiment's lifetime in different orientations on the sky and in many
detectors sensitive to a range of photon wavelengths.

Schematically, we can write the observation as
\begin{equation}
  \label{eq:data}
  d_{\nu t} = \sum_p P_{\nu tp} \Delta_{\nu p} + \eta_{\nu t}  .
\end{equation}
Here, $d_{\nu t}$ is the vector of observations at frequency $\nu$ and
time $t$, $\eta_{\nu t}$ is the noise contribution, $\Delta_{\nu p}$
is the microwave emission at that frequency and position
$p=1,\ldots,m_p$ on the sky, smeared by the experimental beam and
averaged over the pixel.  The pointing matrix, 
$P_{\nu tp}$, is an operator which
describes the location of the beam as a function of
time and its chopping pattern.  For a scanning experiment, it is a
sparse matrix with a 1 whenever position $p$ is observed at time
$t$; for a chopping experiment it will have positive and negative
weights describing the differences made at time $t$. (Note that we
shall often drop the reference to the channel, $\nu$, when referring
to a single frequency).

The first challenge is to separate the noise from the signal and
create an estimate of the map, ${\bar\Delta}_p$, and its noise
properties. This alone is a daunting task: long-term correlations in
the noise mean that the best estimate for the map is not simply a
weighted sum of the observations at that pixel.  Rather, a full
least-squares solution is required. This arises naturally as the
maximum-likelihood estimate of the map if the noise is taken to be
Gaussian (see Eq.~\ref{eqn:pofeta}, below). This in turn requires
complex matrix manipulations due to the long-term noise correlations.

One of the most difficult forms of noise results from the random long
term drifts in the instrument.  These make it hard to measure the
absolute value of temperature on a pixel, though temperature
differences along the path of the beam can be measured quite well
because the drifts are small on short time scales.  However, by the
time the instrument returns to scan a nearby area of the sky, the
offset due to this drift can be quite large, resulting in an apparent
{\em striping} of the sky along the directions of the scan pattern.
The problem is even more complicated than a simple offset because the
detector noise has a ``$1/f$'' component at low frequencies
accompanying the high frequency white noise.

This striping can be reduced by using a better observing strategy.  If
the scan pattern is such that it often passes over one of a set of
well sampled reference points, then the offset can be measured and
removed from the timestreams.  More complicated crossing patterns in
which many pixels are quickly revisited along different scan
directions provide a better sampling of the offset drift and allow it
to be removed more effectively.

The striping issue highlights the global nature of the problem of
map-making.  If the map did not need to be analyzed globally, then one
could cut the map into $N$ pieces and speed up processing time by
$N^2$.  However, including the reference points is essential and these
can be far removed from the subset of pixels in which one is
interested.  More complicated crossing patterns which reduce these
errors unfortunately increase the ``non-locality'' of the problem,
making it difficult to use divide-and-conquer tactics successfully.

Solving for the map in the presence of this noise is, in general, an
$O(m_t^3)$ process, where $m_t$ is the number of elements in the 
time-ordered data.
Since $m_t$ may be anywhere from $10^6$ to upwards of $10^9$, the
general problem cannot be solved in a reasonable time.
Fortunately, the problem becomes tractable if one can exploit the {\it
stationarity}, or time-translation invariance, of the noise.

In addition to solving for the map, one also needs the statistical
properties of the errors in the map.  Accurate calculation of the
``map noise matrix'' is critical, since the signal we are looking for
is excess variance in the map, beyond that which is expected from the
noise.  It turns out that it is both easier to calculate and store the
inverse of the map noise matrix, called the map weight matrix.  The
weight matrix is typically very sparse, whereas its inverse may be
quite dense.  It is therefore advantageous to have algorithms for power
spectrum and parameter estimation which require the weight matrix,
rather than its inverse.

\subsection{Removing the foregrounds\ldots} 

\begin{figure}[htbp]
\begin{center}
\plotonecfpa{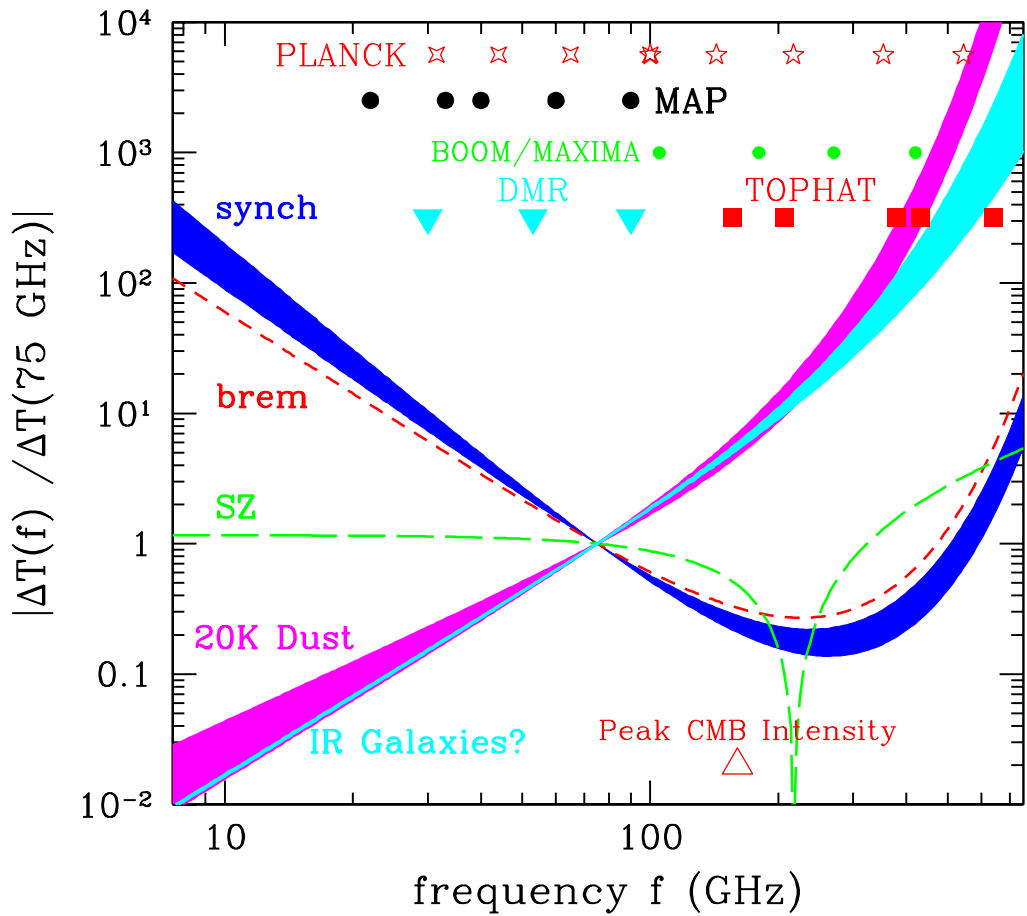}
\caption[]{\baselineskip=10pt\small Foregrounds and secondary
anisotropies depend differently on photon frequency, the key property
for separating the components.  Plotted is the frequency dependence of
the effective thermodynamic temperature fluctuations, normalized to
their values at 75~GHz. Thus the primary CMB fluctuations correspond
to a horizontal line in this figure. At long wavelengths, synchrotron
and bremsstrahlung are troublesome, while at shorter wavelengths dust
is a problem. The bands represent a measure of our uncertainty in the
appropriate foreground shapes.  The highly distinctive shape for the
Sunyaev-Zeldovich (SZ) effect is negative at low frequencies, positive at
high. The actual level of contamination depends on the angular scale;
{\it e.g.}, one can estimate the values for dust and synchrotron
emitting radio sources by multiplying these curves by the appropriate
values from Figure~\ref{fig:CLtheory}.  Most CMB experiments take
measurements at a number of frequencies centered around 90~GHz, where
the foreground emission is minimal, and not far from where the CMB
{\em intensity} peaks. Detector frequencies for some notable
experiments, in particular those of Fig.~\ref{fig:clproj}, are denoted
by the symbols at the top.  Notice the wide coverage planned for the
Planck satellite. }
\label{fig:frequencyforegrounds}
\end{center}
\end{figure}

Maps are made at a number of different wavelengths.  Each of these
maps will be the sum of the CMB signal, $T_p$, and contributions from
astrophysical foregrounds: sources of microwave emission in the
universe other than the CMB itself.  This includes low-frequency
galactic emission from the 20K dust that permeates the galaxy and from
gas, emitting synchrotron and bremsstrahlung (or free-free)
radiation. There are also extragalactic sources of emission: galaxies
that emit in the infrared and the radio. These are treated as
point sources, since their
angular size is much smaller than the experimental
resolution. In addition, clusters of galaxies and the filamentary
structures connecting them will appear because their hot gas of
electrons can Compton scatter CMB photons to shorter wavelengths, a
phenomenon known as the Sunyaev-Zel'dovich (SZ) effect. These clusters are
typically a few arcminutes across, small enough to be resolved by
Planck but not MAP.  In Figure~\ref{fig:spatialforegrounds}, we
schematically show the spatial patterns of some of these foregrounds,
and in Figure~\ref{fig:frequencyforegrounds}, we show their frequency
spectra.

The next challenge, then, is to separate these foregrounds from the CMB
itself in the noisy maps. We write 
\begin{equation}
  \label{eq:noisyfgmaps}
  {\bar\Delta}_{\nu p} = T_p + \sum_i f^{(i)}_{\nu p} + n_{\nu p}.
\end{equation}
Here, $T$ is the frequency-independent CMB temperature fluctuation, $n$
is the noise contribution whose statistics have been calculated in the
map-making procedure, and $f^{(i)}_{\nu p}$ is the contribution of the
foreground or secondary anisotropy component $i$. The shapes of the
expected frequency dependences shown in
Figure~\ref{fig:frequencyforegrounds} show some uncertainty. There is
none for some secondary anisotropy sources, {\it e.g.}, the
Sunyaev-Zeldovich effect, so $ f^{(i)}_{\nu p}$ can be considered a
product of the given function of frequency times a spatial function.  In
the past, an approximation like this involving a single spatial template
and one function of frequency has been used for all of the foregrounds,
but it is essential to consider fluctuations about this for the accuracy
that will be needed in the data sets to come.

A crude but reasonably effective method is to separate the signals
using the multifrequency data on a pixel-by-pixel basis. However, it is
clearly better to use our knowledge of the spatial patterns in the
forms adopted for ${\cal P}(f^{(i)}_{\nu p}\vert {\rm theory})$, {\it
e.g.}, the foreground power spectra shown in
Fig.~\ref{fig:CLtheory}. Even using a Gaussian approximation for the
foreground prior probabilities has been shown to be relatively
effective at recovering the signals. In this case, the statistical
distribution of the maps is again Gaussian, with a mean given by the
maximum likelihood, which turns out to involve {\it Wiener
filtering} of the data [\cite{numrec}]. 
In simulations for Planck performed by Bouchet and Gispert,
the layers making up the ``cosmic sandwich'' in
figure~\ref{fig:spatialforegrounds} have been convolved with the
frequency-dependent beams, and realistic noise has been added.  The
recovered signals look remarkably like the input ones. There is some
indication that the performance degrades if too large a patch of the
sky is taken, possibly because the non-Gaussian aspects become more
important. Of course, good estimates of the power spectra for each of
the foregrounds are essential ingredients for ${\cal P}(f^{(i)}_{\nu
p}\vert {\rm theory})$, and these must be obtained from the CMB data in
question by iterative techniques, or with other CMB data.

Radio astronomers have a long history of image construction using
interferometry data. One of the most effective techniques is the
``maximum entropy method''. Although this is often a catch-all phrase for
finding the maximum likelihood solution, the implementation of the
method involves a specific assumption for the nature of ${\cal
P}(f^{(i)}_{\nu p}\vert {\rm theory})$, derived as a limit of a Poisson
distribution. For small fluctuations it looks like a Gaussian, but has
higher probability in the tails than the Gaussian does. The Poisson
aspect makes it well-suited to find and reconstruct point
sources. To apply it to the CMB, which has both positive and negative
excursions, and to include signal correlation function information,
some development of the approach was needed. This has been recently 
carried out and applied to 
the cosmic sandwich exercise [\cite{bouch}]. 
It did at least as well at recovery as
the Wiener method did, and was superior for the concentrated
Sunyaev-Zeldovich cluster sources and more generally for point
sources, as might be expected. Errors on the maximum entropy maps are
estimated from the second derivative matrix of the likelihood
function. 

We regard these exercises as highly encouraging, but since the
accuracy with which cosmological parameters can be determined is
very dependent upon the accuracy with which separation can be
done, it is clear that much work is in order for improving the
separation algorithms.

\subsection{From the CMB to cosmology\ldots} 
Armed with a CMB map and its noise properties, we can try to extract
its cosmological information.  If we assume the cosmological signal is
the result of a statistically isotropic Gaussian random process, then
all of the information is contained in the power spectrum, $C_\ell$.
With Gaussian noise as well, we can write down the exact form of its
likelihood function.  Unfortunately, because of incomplete sky
coverage, and the presence of correlated, anisotropic noise,
maximizing this likelihood function (either directly or by some sort
of an iterative procedure) requires manipulation of $m_p\times m_p$
matrices, typically needing $O(m_p^3)$ operations and $O(m_p^2)$
storage. This becomes computationally prohibitive on typical
workstations when $m_p$ exceeds about $10^4$; for the $m_p>10^6$
satellite missions even supercomputers may be inadequate to the
task. For example, on a single 1000~MHz processor, even one
calculation of $O(10^{21})$ operations necessary for a
ten-million-pixel map would take {\em 30,000~years}\/!
There is, as of yet, no
general solution to this problem.  However, in some cases, such as for
the MAP satellite, a solution has been proposed which relies upon the
statistical isotropy of the signal and a simple form for the noise.
Unfortunately, most experiments will produce maps with more complicated
noise properties.

The power spectrum is a highly compressed form of the data in the map,
but it is not the end of the story.  The real goal remains to
determine the underlying {\em cosmological parameters}, such as the
density of the different components in the universe.  For the simple
inflationary models usually considered, there are still at least ten
different parameters which affect the CMB power spectrum, so we must
find the best fit in a ten (or more) dimensional parameter space. Just
as the frequency channel maps were derived from the timestreams, the
CMB map from the frequency maps, and the power spectrum from the CMB
map, the cosmological parameters can be estimated from the power
spectrum.  Although in doing so, one must be careful about the
non-Gaussian distribution of the uncertainty in the $C_\ell$
[\cite{bjkII}].

\section{The Most Daunting Challenges}

We now take a more in-depth look at the problems of map-making and
parameter estimation.  The most general algorithms for solving these
problems operate globally on the data set and are prohibitively
expensive: both require matrix operations $O(m^3)$, where $m$
is either the number of points in the time series ($m_t>10^9$ for
upcoming satellites) or the number of pixels on the sky
($m_p>10^6$). Special properties, such as the approximate {\it
stationarity}\/ of the instrumental noise, must be exploited in order
to make the analysis of large data sets possible.  To date
most work has concentrated on efficient algorithms for the exact
global problem, but for the new data sets it will be essential to develop
approximate methods as well.

We wish to find the {\it most likely}\/ maps and power spectra.  We
can write down likelihood functions for both these quantities if we
assume that both the noise and signal are Gaussian.  While the
maximum-likelihood map has a closed-form solution, there is no such
solution for the most likely power spectrum.  Thus, the problem of the
cost of evaluating the likelihood function is compounded by having to
search a very high-dimensional space for the global maximum.

Even these complex problems are an oversimplification because we know
that foregrounds and secondary anisotropies have non-Gaussian
distributions.  Thus, although we expect to get valuable results using
simplified approximations for ${\cal P}(f^{(i)}_{\nu p} \vert {\rm
th})$, in particular the Gaussian one we use in the discussion below,
Monte Carlo approaches in which many $f^{(i)}_{\nu p}$ maps are made
will undoubtedly be necessary to accurately determine the uncertainty
in the derived cosmological parameter.

\subsection{Map-making:  the ideal case}
As described in Eq.~\ref{eq:data}, for each channel we 
model the timestream, $d$, as due to signal,
$\Delta$, and noise, $\eta$, $d = P\Delta + \eta$, where $P$
is the pointing matrix that describes the observing strategy as a
function of time.  In the ideal case, the noise is
Gaussian-distributed, {\it i.e.}, its probability distribution is 
\be
\label{eqn:pofeta}
{\cal P}(\eta) = \left[ \left(2\pi\right)^{m_t} |N| \right]^{-1/2}
\exp\left(-\eta^\dagger N^{-1} \eta/2\right) \, , 
\ee 
where $m_t$ is the number of time-ordered data points and $N_{tt'} \equiv
\langle \eta_t \eta_{t'}^\dagger \rangle$ is the noise covariance
matrix. Here the $^\dagger$ denotes transpose and the brackets
indicate an ensemble average (integration over ${\cal P}(\eta)
d\eta$).  Substituting $d-P\Delta$ for $\eta$ in this expression gives
the probability of the time-ordered data given a map, ${\cal
P}(d|\Delta)$, which is also referred to as the likelihood of the map,
${\cal L}(\Delta)$.  We are actually interested in the probability
of a map given the data, ${\cal P}(\Delta|d)$.  If we assign a uniform
prior probability to the underlying map, {\it i.e.}, ${\cal P}(\Delta
\vert {\rm theory} )$ is constant, then by Bayes' theorem ${\cal
P}(\Delta|d)$ is simply proportional to the likelihood function, 
${\cal L}(\Delta)$.

The map that maximizes this likelihood function is
\be
\label{eqn:mapsoln}
{\bar\Delta} = C_N P^\dagger N^{-1} d
\ee
where $C_N$ is the noise covariance matrix of the map,
\be
C_N\equiv \langle \left({\bar\Delta}-\Delta\right)
\left({\bar\Delta}-\Delta\right)^\dagger \rangle =
\left(P^\dagger N^{-1} P\right)^{-1}.
\ee
This map is known as a {\em sufficient statistic}, in that
$\bar{\Delta}$ and $C_N$ contain all of the sky information in the
original data set, provided the pixels are small enough.  As discussed
above, it is preferable to work with $C_N^{-1}$, the map weight
matrix, which is often sparse or nearly so.

For many purposes, the variance-weighted map, 
\be
\label{eqn:solveforDelta}
C_N^{-1}{\bar \Delta} = P^\dagger N^{-1}d
\ee
may be more useful than the map itself, so that we can avoid the
computationally intensive step of inverting the weight matrix.  This
is true for optimally combining maps, since variance-weighted maps and
their weight matrices simply sum, and for finding the minimum-variance
map in a different basis, such as Fourier modes or spherical
harmonics.  An algorithm for finding the most likely power spectrum
exploits this, as we will see below.

If we do need to find ${\bar \Delta}$, we can solve 
Eq.~\ref{eqn:solveforDelta} iteratively by techniques
like the conjugate gradient method.  In general, such methods
require $m_p$ iterations and are effectively still
$O(m_p^3)$ methods.  Fortunately, we
expect $C_N$ to be sufficiently diagonal-dominant that
many fewer than $m_p$ iterations are required.  This is aided by
the use of {\it pre-conditioners}, which will be discussed
further in the context of finding the maximum-likelihood power spectrum.

Whether we are interested in ${\bar \Delta}$ or $C_N^{-1}{\bar
\Delta}$, we still must convolve the inverse of $N$ with the data
vector.  The direct inversion of $N$ by brute force is impractical
since it is an $m_t \times m_t$ matrix where $m_t$ is often about
$10^9$.  However, this is greatly simplified if the noise is
stationary, which means its statistical properties are time
translation invariant, so that $N_{tt'}=N(t-t')$.  Stationarity 
means that $N$ is diagonal in Fourier space with eigenvalues
$\widetilde{N}(f)$, the noise {\em power spectrum}.  $N^{-1}$ is then
just the inverse Fourier Transform of $1/\widetilde{N}(f)$.
Knowing $N^{-1}$, it is easy to calculate
the map weight matrix, $C_N^{-1} = P^\dagger N^{-1} P$.
 
The convolution of $N^{-1}$ with $d$ appears to be an $O(m_t^2)$
operation.  Since there is much more timestream data $(m_t \gg m_p)$,
this is potentially the slowest step in the calculation of the map.
Fortunately, the convolution is actually much faster because
$N^{-1}(t-t')$ generally goes nearly to zero for $t-t' \gg 0$.
The absence
of weight at long time scales can be due to the ``$1/f$'' nature of the
instrument noise at low temporal frequencies. 
Atmospheric fluctuations
also have more power on long time scales than on short time scales, as
do many noise sources.  Since these characteristic times do  
not scale with the mission duration, the convolution is actually
$O(m_t)$.  Similarly, the multiplication of the pointing matrix is
also $O(m_t)$ because of its sparseness.

Thus, we can reduce the timestream data to an estimate of the map
and its weight matrix in only
$O(m_p^2)$ operations, a substantial savings compared to
the $O(m_t^3)$ operations required for a direct calculation.
These algorithms, or similar ones, have been
implemented in practice, {\it e.g.}, [\cite{qmap,wri}].

\subsection{Map-making:  complications}

Above, we made two simplifying assumptions: that the statistical
properties of the noise in the timestream were known and that the noise
sources were all stationary.  Here we try to deal with the more
general case.

We would like to estimate the statistical properties of the noise by
using a model of the instrument, but in practice, these models are
never sufficient.  One must always estimate the noise from the data set
itself, and doing this from the timestream requires some assumptions.
It is usually assumed that the noise is stationary over sufficiently
long intervals of time and is Gaussian.  Often the data set is
dominated by noise and to a first approximation, is all noise.  Thus
one has many pairs of points separated by $t-t'$ to estimate
$N(|t-t'|) = \langle \eta(t)\eta(t') \rangle$.
Techniques are being developed [\cite{fjnoise}] to
simultaneously determine the map and noise power spectrum and the
covariance between the two.

Non-stationary noise can arise in a number of ways: possible sources
include contamination by radiation from the ground, balloon or sun,
some components of atmospheric fluctuations and cosmic ray hits.  
Often they are
synchronous with a periodic motion of the instrument.  They can be
taken into account by extending the model of the timestream given in
Eq.~\ref{eq:data} to include contaminants of amplitude $\kappa_c$ with
a known ``timestream shape'', $\Upsilon_{tc}$:
\begin{equation}
  d_t = \sum_p P_{tp} \Delta_p + \sum_c \Upsilon_{tc} \kappa_c +\eta_t.
\end{equation}
The contaminant amplitudes are now on the same mathematical
footing as the map pixels, $\Delta_p$, and both can be solved for
simultaneously.

A more conservative approach assigns infinite noise to modes 
of the time-ordered data which can be written as a linear combination of the
$\Upsilon_{tc}$.  Doing so removes all sensitivity of the map to the
contaminant, irrespective of the assumption of Gaussianity.
Operationally, we replace the timestream noise covariance matrix, $N_{tt'}$ with 
\be
\label{eq:matcon} 
N_{tt'}
\rightarrow N_{tt'}+\sum_c \sigma^2_c \Upsilon_{tc} \Upsilon_{t'c} 
\ee
where the $\sigma_c^2$ are taken to be very large, thereby setting
the appropriate eigenvalues of $N^{-1}$ to zero. 

This noise matrix has lost its time-translation invariance
and so is no longer directly invertible by Fourier transform methods.
Fortunately, there is a theorem called 
the {\it Woodbury Formula} [\cite{numrec}] which allows one to find
the resulting correction to $N^{-1}$ for additions to $N$ of the
form in Eq.~\ref{eq:matcon} while only having to
invert matrices of dimension equal to the number of contaminants.

\subsection{Parameter Estimation: A First Attempt }\label{sec:paramest}

We now turn to the determination of some set of cosmological
parameters from the map. We will focus on the case where the
parameters are the $C_\ell$'s because it is a model independent way of
compressing the data.  However, the discussion below can easily be
generalized to any kind of parameterization, including the ten or more
cosmological parameters that we would like to constrain.

We wish to evaluate the likelihood of the parameters ${\cal
L}(C_\ell)\equiv {\cal P}({\bar T} | C_\ell)$, which folds in the
probability of the map given the data with all of the prior
probability distributions, for the target signal $T$ and the
foregrounds and secondary anisotropies $f^{(i)}$, in a Bayesian way:
\begin{equation}
{\cal L}(C_\ell) = \int {\cal
P}(\Delta|d) \, {\cal P} (T| {\rm theory})d^{m_p}T \, \prod_i {\cal P}
(f^{(i)}|{\rm theory})d^{m_p}f^{(i)}\, .  
\end{equation}
Only in the Gaussian or uniform prior cases is the integration over
$T$ and $f^{(i)}$ analytically calculable. The usual procedure for ``maximum
entropy'' priors is to estimate errors from the second derivative of
the likelihood, {\it i.e.} effectively use a Gaussian
approximation. Exploring how to break away from the Gaussian
assumption is an important research topic. 

Assuming all signals and the noise are Gaussian-distributed, the
likelihood function is
\begin{equation}
  \label{eq:like}
{\cal L}(C_\ell)  =  
{  \exp\left[
    -{1\over2} {\bar T}^\dagger\left(C_N + C_S\right)^{-1}{\bar T}\right]
\over
\left[ \left(2\pi\right)^{m_p} |C_N + C_S| \right]^{1/2}
},
\end{equation}
where $\bar T$ is the maximum-likelihood CMB map, with the 
foregrounds removed. 
$C_N$ is the noise matrix 
calculated above, modified to include variances determined for the 
foreground maps, and $C_S$ is the primary signal autocorrelation
function which depends on $C_\ell$
(as in Eq.~\ref{eqn:TT}, but corrected for the effect of the 
beam pattern and finite pixel size).

The likelihood function is a Gaussian distribution {\em in the data},
but a complicated nonlinear function of the parameters, which enter
into $C_S$ through the power spectrum.  Unlike the map-making problem
(Eq.~\ref{eqn:mapsoln}), there is no closed-form solution for the most 
likely $C_{\ell}$.  Thus we must use a search strategy and it
should be a very efficient one,  since brute force evaluation of the
likelihood function requires determinant evaluation and matrix
inversion which are both $O(m_p^3)$ problems.  
Compounding this, evaluating the likelihood is more difficult here because 
the signal and noise matrices have different symmetries, making it 
harder to find a basis in which $C \equiv C_S + C_N$ has a simple form. 

A particularly efficient search technique for finding the
maximum-likelihood parameters is a
generalization of the {\em Newton-Raphson}\/ method of root finding.
The Newton-Raphson method finds the zero of a function of one
parameter iteratively.  One guesses a solution and corrects that guess
based on the first derivative of the function at that point. 
If the function is linear, this correction is exact; otherwise, 
more iterations are required until it converges. 

In maximizing the likelihood, we are searching for regions where the
first derivative of the likelihood with respect to the parameters goes
through zero, so it can be solved analogously to the Newton-Raphson
method.  We actually maximize $\ln{\cal L}$, which simplifies the
calculation and also speeds its convergence since the derivative of
the logarithm is generally much more linear in $C_\ell$ than the
derivative of the likelihood itself.  Solving for the roots of
$\partial \ln{\cal L}/\partial C_\ell$ using the Newton-Raphson method
requires that we calculate $\partial^2 \ln{\cal L}/\partial C_\ell \partial
C_{\ell'}$, which is known as the curvature of the likelihood
function.  Operationally, we often replace the curvature with its
expectation value $F_{\ell\ell'}$, the {\em Fisher matrix}, because it
is easier to calculate and still results in convergence to the same
parameters.

The change in the parameter values at each iteration for this method
is a quadratic form involving the map; hence it is referred to as a
{\em quadratic estimator}.  Using $C_\ell$ as our parameter, the new
guess is modified by [\cite{bjkI,teg}]
\begin{equation}
  \label{eq:quadraticCl}
  \delta C_\ell =  {1\over 2}\sum_{\ell'} F^{-1}_{\ell\ell'}
  \left[{\bar T}^\dagger  C^{-1} {\partial C \over \partial C_\ell} C^{-1} {\bar T} 
-{\rm Tr}\left(  
      {\partial C \over \partial C_\ell} C^{-1}\right)\right]
\end{equation}
where the Fisher matrix is given by 
\begin{equation}
  \label{eq:fish}
  F_{\ell\ell'} \equiv \left\langle -{\partial^2 \ln {\cal L}\over \partial
C_\ell \partial C_{\ell'}}\right\rangle
=  {1\over2}{\rm Tr}\left( C^{-1}{\partial C \over \partial C_\ell} 
C^{-1} {\partial C \over \partial C_\ell} \right) \, .
\end{equation}
We can recover the full shape of the likelihood
for the $C_\ell$'s from this and one other set of numbers, calculated
in approximately the same number of steps as the Fisher matrix itself
[\cite{bjkII}]. 

The procedure is very similar to that of the Levenberg-Marquardt
method [\cite{numrec}] for minimizing a $\chi^2$ with non-linear
parameter dependence.  There the curvature matrix (second derivative
of the $\chi^2$) is replaced by its expectation value and then scaled
according to whether the $\chi^2$ is reduced or increased from the
previous iteration.  Similar manipulations
may possibly speed convergence of the likelihood maximization,
although one would want to do this without direct evaluation of the
likelihood function.

This method has been used for the power spectrum estimates for COBE
and other experiments, and for the compressed power spectrum bands
estimated from current data shown in Fig.~\ref{fig:CLtheory}.  This
brute force approach is quite tractable for the current data and for
idealized simulations of the satellite and LDB data, such as the power
spectrum forecasts of Fig.~\ref{fig:clproj}, in which the noise was
assumed (incorrectly) to be homogeneous.

We can calculate the time and memory required to do this quadratic
estimation for a variety of realistic data sets and kinds of computing
hardware.  For this algorithm, the $O(m_p^3)$ operations must be
performed for each parameter (e.g., each band of $\ell$ for $C_\ell$).
Borrill [\cite{borr}] has considered this issue under several
different scenarios. For COBE, power spectrum calculation can easily
be done on a modern workstation in less than one day.  However, for
the LDB data sets expected over the next several years (with $m_p \ga
200,000$ or so) the required computing power becomes prohibitive,
requiring 640~Gb of memory and of order $3\times10^{17}$
floating-point operations, which translates to {\em 40~years}\/ of
computer time at 400~MHz.
This pushes the limits of available technology; even spread over a
Cray T3E with $\sim1024$ 900~MHz processors, this would take a week or
more.  This data set is in hand {\em now}, so we cannot even wait for
computers to speed up.  When the satellite data arrives, with
$m_p>10^6$, a brute-force calculation will clearly be impossible even
with projected advances in computing technology over the next
decade. The ten million pixel Planck data set would require 1600~TB of
storage and $3\times10^{23}$ floating-point operations or 25,000 years
of serial CPU time at 400~MHz.  Even a hundredfold increase in
computing over the next decade, predicted by Moore's law, still
renders this infeasible.

\subsection{Discretizing the Sky}
\label{sec:pixel}

To solve these computing challenges, shortcuts must be found. 
One area where there is great potential benefit is in deciding 
how the discretized map elements are to be distributed on the sky and stored.  
Imposing enough symmetries at this early step can help greatly to speed 
up everything that follows. 

Obviously it is important to keep the number of pixels as small as possible. 
For a given resolution, fixed for example by the beam size, the 
number of pixels is minimized by having them all roughly of the same area.  
If there are many pixels in a resolution element much smaller than the beam size, 
they will be highly correlated and little information is gained by treating them 
individually.

The hierarchical nature of
the pixelization used for the COBE maps was also a very useful property. 
In this pixelization, known as the Quadrilateralized 
Spherical Cube, the sky was broken into six base pixels corresponding 
to faces of a cube.  Higher resolution pixels were created hierarchically, 
by dividing each pixel into four smaller pixels of approximately equal area.  
One advantage of this hierarchical structure is that the data  is effectively 
stored via a branching structure, so that pixels that are physically close to 
each other are stored close to each other.  Among other things, this allows one to 
coarsen a map very quickly, by adding the ordered pixels in groups of four. 

Finally, it is very beneficial to have a pixelization which is azimuthal, 
where many pixels share a common latitude.  This is incredibly useful in 
making spherical harmonic transforms between the pixel space, where the data    
and inverse noise matrix are simply defined, and multipole space, where the theories 
are simple to describe.   Specifically, one wishes to make transforms of the type 
described by Eq.~\ref{eqn:sph-trn}, as well as the inverse transformation.
When discretized, these transforms naively take $m_p^2$ operations,
because $m_p$ spherical harmonic functions need to be evaluated
at $m_p$ separate points on the sky.  

However, as has been recently emphasized, if one uses a pixelization
with  azimuthal symmetry, then the spherical transforms can be greatly
sped up [\cite{mnv}].
This utilizes the fact that the azimuthal dependence of the
spherical harmonic functions can be simply factored out,
$ Y_{\ell m}(\theta,\phi) = \lambda_{\ell m}(\theta) e^{im\phi}$. 
If one further requires that the pixels have discrete azimuthal 
symmetry, then the azimuthal sum can be performed quickly with 
a fast Fourier transform.
Effectively, this means that the $m_p$ functions need only be evaluated
at $m_p^{1/2}$ different latitudes, so that the whole process requires
only $m_p^{3/2}$ operations.
Efforts have been made to speed this up even further, by attempting to use 
FFT's in the $\theta$ direction as well, which in principle could perform 
the transform in $m_p (\log m_p)^2$ operations.  Such implementations  
are still being developed, and do not tend to pay off until $m_p$ is 
very large. 

Pixelizations have been developed which have all of these symmetries.
HEALPix, devised by Kris Gorski and collaborators [\cite{ghw}], has a
rhombic dodecahedron as its fundamental base, which can be divided
hierarchally while remaining azimuthal.  It was used for the rapid
construction of the map in Fig.~\ref{fig:allsky}. Another class of
pixelizations is based on a naturally azimuthal igloo structure which
has been specially designed to be hierarchical [\cite{ct}].  In this
scheme, pixel edges lie along lines of constant latitude and
longitude, so it is easy to integrate over each pixel exactly. This
allows any suppression effects due to averaging over the varying pixel
shapes to be simply and accurately included when making the
transforms.

\subsection{Exploiting the Symmetries}\label{sec:osh}

Since many of the signals are most simply described in multipole
space, it is natural to try to exploit this basis when implementing the
parameter estimation method described above. 
We should also try recasting the calculation to take advantage of the simple 
form the weight matrix $C_N^{-1}$ has in the pixel basis.  
Finally, with iterative methods 
we can exploit approximate symmetries of these matrices 
which can speed up the algorithms tremendously. 
Oh, Spergel and Hinshaw [\cite{osh}], hereafter OSH, have recently
applied these techniques to simulations of the operation of parameter
estimation for the MAP satellite to great effect.

The Newton-Raphson method does not require
the full inverse correlation matrix, but rather $C^{-1} {\bar T}$,
which can be expressed in terms of $C_N^{-1}$ and various $C_S^{1/2}$
factors.  The equation can be solved using a simple conjugate gradient
technique, which iteratively solves the linear system $ C z = {\bar
T}$ by generating an improved guess and a new search direction
(orthogonal to previous search directions) at each step.  In general,
conjugate gradient is no faster than ordinary methods, requiring of
order $m_p$ iterations with $m_p^2$ operations per iteration required
for the matrix-vector multiplications.  However, this can be sped up in two
ways. First, one can make the matrix well conditioned by finding an
appropriate preconditioner which allows the series to converge much
faster, in only a few iterations.  Second, one can exploit whatever
symmetries exist to do the multiplications in fewer operations.

A preconditioner $\widetilde{C}$ is a matrix which approximately
solves the linear system and is used to transform it to
$\widetilde{C}^{-1}C z = \widetilde{C}^{-1} {\bar T}$, making the
series converge much faster.  There are two requirements of
a good preconditioner: it should be close enough to the original
matrix to be useful and it should be quickly invertible.  One
can rewrite the linear system we need to solve as
\be \left(I +
C_S^{1/2}C_N^{-1}C_S^{1/2}\right) C_S^{1/2}z = C_S^{1/2}C_N^{-1} {\bar
T} .    
\ee
OSH use a preconditioner $\left(I +
C_S^{1/2}\widetilde{C}_N^{-1}C_S^{1/2}\right)$, where
$\widetilde{C}_N^{-1}$ is an approximation to the inverse noise matrix
in multipole space: $\widetilde{C}_N^{-1}$ is taken to be azimuthally
symmetric, so that it is proportional to $\delta_{mm'}$ in multipole
space, which makes it block diagonal and possible to invert quickly.
For the case they looked at, which includes only uncorrelated
pixel noise and an azimuthally symmetric sky cut, this turned out to be
a very good approximation which allows for quick convergence.

Because the matrices are simple in the bases chosen, the vector-matrix
multiplications are much faster than $m_p^2$ .  In multipole space,
the theory correlation matrix is simply diagonal, $C_S = C_\ell
B_\ell^2 \delta_{\ell\ell'} \delta_{mm'}$, where $B_\ell$ denotes the
beam pattern in $\ell$ space.  Similarly, in pixel space, operations
using the inverse noise matrix are much faster.  (OSH simplified
to a case where the noise matrix was exactly diagonal in pixel space.)
A time-consuming aspect is the transformation between pixel and
multipole space, which is $O(m_p^{3/2})$.  The whole process
is actually dominated by the calculation of the trace in
Eq.~\ref{eq:quadraticCl}, which is performed by Monte Carlo iterations
of the above method, exploiting the fact that $\langle{\bar T}^\dagger
C^{-1} \partial C/ \partial C_\ell C^{-1} {\bar T} \rangle = 
{\rm{Tr}}[\partial C_S/ \partial C_{\ell'} C^{-1}]$.  
The OSH method requires effectively $m_p^2$ operations, a
dramatic improvement over traditional methods.

\section{Unsolved Problems} 

The methods highlighted here have focused on solving one 
well-posed problem under a number of important simplifying assumptions.  It
is not obvious whether any of these assumptions are correct or
indeed if the problem itself is as simple as we have described.  In
addition, there remain other problems, as or more complex, which
remain to be addressed.  Here, we briefly touch on some of these
issues.

The improvements in speed discussed in the last section relied heavily
on assuming the error matrix was close to being both diagonal and
azimuthally symmetric.  This may well be the case for the MAP
satellite, because it measures the temperature difference between each
point on the sky and very many other points at a fixed angular
separation of $120^\circ$ at many different time scales.  In doing so,
the off-diagonal elements of the noise matrix are ``beaten down'' and
may indeed be negligible.  However, for almost all other cases (and
indeed possibly for MAP when the effects of foreground subtraction are
taken into account,) the $C_\ell$ estimation problem becomes much more
complicated.  In the presence of significant striping or inhomogeneous
sky coverage, the block-diagonality of the noise matrix is no longer a
good approximation.  In this case, finding a basis where both the
signal and noise matrices are simple may not be possible.  People have
found signal-to-noise eigenmodes of the matrix
$C_N^{-1/2}C_SC_N^{-1/2}$ (or $C_S^{1/2}C_N^{-1}C_S^{1/2}$ as in
Sec.~\ref{sec:osh}) to be useful for data compression and computation
speedup, but finding them is another $O(m_p^3)$ problem.

One might try to solve this by splitting the data set up into smaller
bits and analyzing them separately, recombining the results at the
end.  However, as emphasized above, this can be difficult to do
because of the global nature of the the mapmaking process.  Ignoring
correlations between different regions is often a poor approximation.
Due to the complicated noise correlation structure, optimally
splitting and recombining may itself require the $O(m_p^3)$ operations
we are trying to avoid.

Another feature of realistic experiments that has not been properly
accounted for in the formalism we have outlined is that of asymmetric
or time-varying beams. The model of the experimental procedure we have
given here (Eq.~\ref{eq:data}) assumes that all observations of a
given pixel see the same temperature. This implicitly assumes an
underlying model of the sky that has been both beam-smoothed and
pixelized. (Pixelization effects were touched on in
Sec.~\ref{sec:pixel}.) If the beam is not symmetric, or if it is
time-varying, then different sweeps through the same pixel will see
different sky temperatures.  This is very difficult to account for
exactly and may be crucial for some upcoming experiments which can
have significantly asymmetric beams.

In addition, large uncertainties in the nature of the foregrounds may
make their removal quite tricky.  Not only are they non-Gaussian, but
unlike the CMB, their frequency dependence is not well understood.
Above, we have cast the problem of foreground separation as
essentially a separate step in the process, between the making of maps
at various frequencies and the estimation of the cosmological power
spectrum.  However, we may need to study foregrounds contaminants in
as much detail as the CMB fluctuations themselves in order to fully
understand their impact on parameter determination.

Throughout, we have emphasized the assumption of Gaussianity for both
the instrumental noise and the cosmological model. If one or both of
these assumptions are violated, the theoretical underpinning of the
algorithms we have described becomes shaky.  Non-Gaussianity issues
arise even in intrinsically Gaussian theories, due to foregrounds and
non-linear effects.  More worrisome are models with intrinsic
non-Gaussianity at larger angular scales.

How do we even begin to characterize an arbitrary distribution of sky
temperatures? As it is sometimes put, describing non-Gaussian
distributions is like describing ``non-dog animals.''  However,
techniques do exist for finding specific flavors of non-Gaussianity;
for example, estimations have been made recently of the so-called
connected $n$-point functions for $n>2$ which vanish for a Gaussian
theory.  Other methods have tried to find structures using wavelets,
which localize phenomena in both position on the sky
and scale (wavenumber $\ell$).  Still others have attempted to find
topological measures of non-Gaussianity, focusing on fixed temperature
contours, like the isotherms of a weather map.  For all of these
cases, however, both the theoretical predictions and data analysis are
considerably more difficult than the algorithms presented here; in
particular, none of them have been considered in the presence of
complicated correlated noise.

The computational challenges we have highlighted are associated
specifically with parameter estimation from CMB data, but the problems
are generic to other statistical measures that might be of interest.
For example, goodness-of-fit tests (like a simple $\chi^2$ or more
complicated examples like those explored in [\cite{qmap,knoxcompare}])
require calculation of a quadratic form involving inversion of
$m_p\times m_p$ matrices, as in the parameter estimation examples
above.  One might hope that these problems may also be solvable given
similar assumptions to those considered above, but this has yet to be
addressed.

Finally, we have not even touched on the problem of analyzing
measurements of the polarization of the CMB, which results from
Thomson scattering at the surface of last scattering. Although the
essential aspects of the analysis are the same, polarization data will
be considerably more difficult to handle for several reasons.
First, because polarization is defined with respect to spatially fixed
axes, we must combine measurements from different experimental
channels in order to make an appropriate sky map.  Second, the signal
is expected to be about one tenth the amplitude of the already very small
temperature anisotropies.  Third, the polarization of foreground
contaminants is even less well-understood than their temperatures.
With these greater experimental
challenges, the resulting maps, and their construction algorithms,
will be more complicated.

\section{Finale} 

Upcoming CMB data sets will contain within them many of the answers to
questions that have interested cosmologist for decades: How much
matter is there in the universe? What does it consist of? What did the
universe look like at very early times? Our task will be to extract
the answers and assess the errors from these large
data sets. Especially challenging are the necessities for a {\em
global}\/ analysis of the data and for separating the various
signals. Although some of the issues we face are specific to the CMB
problem, many are of common concern to all astronomers facing the huge
onslaught of data from the ground, balloons and space that the next
millennium is bringing (see, {\it e.g.}  the article on the Sloan
Digital Sky Survey). We cannot rely on raw computing power
alone. 
Computer scientists and statisticians are now collaborating
with cosmologists in the quest for algorithmic advances.
\acknowledgements

\begin{figure}
\begin{center}
\plotone{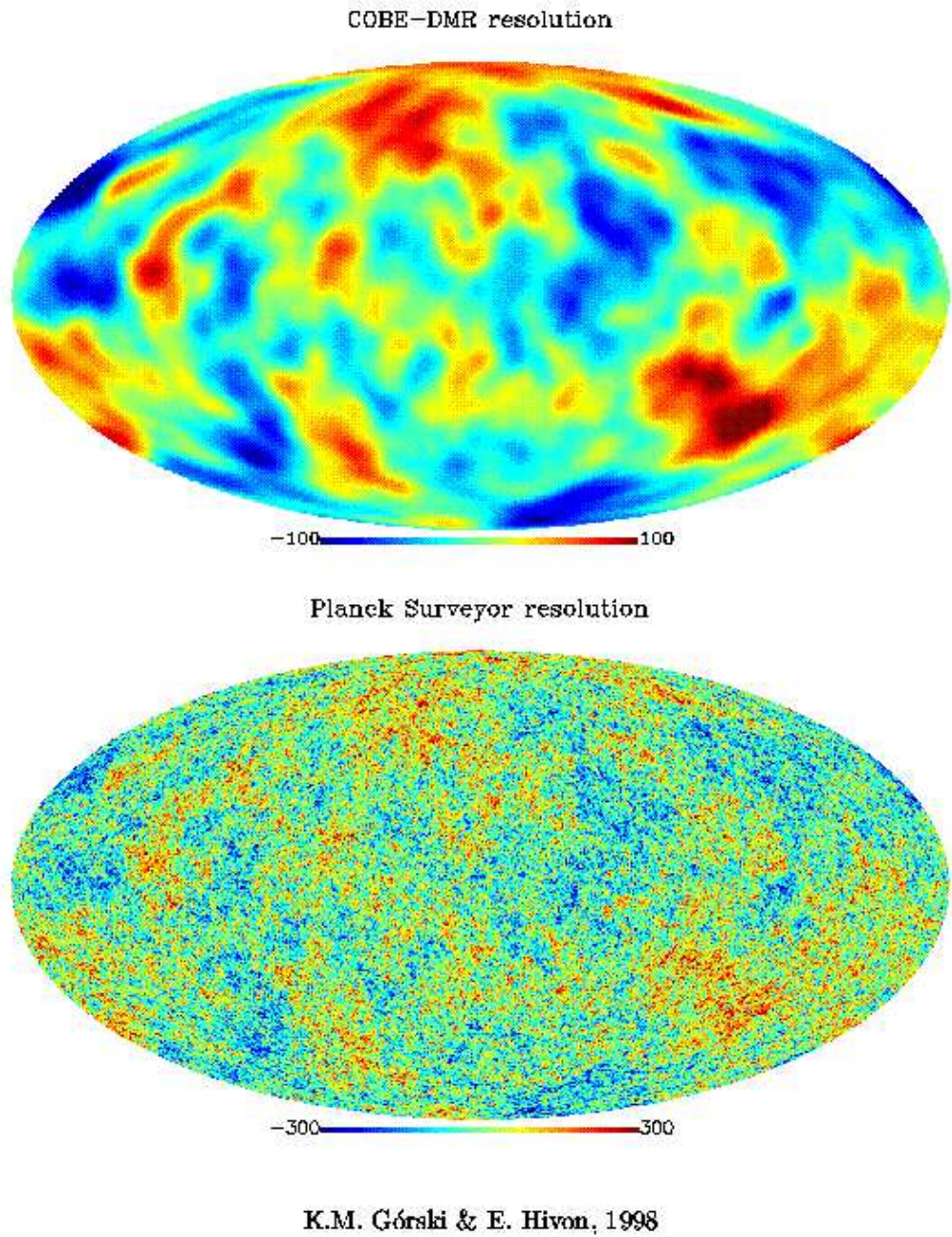}
\caption[ALLsky]{\baselineskip=10pt\small 
A simulated all-sky map of
Gaussian-distributed microwave background temperature
fluctuations. The upper map is at the resolution of the COBE satellite
($\sim 7^\circ$ full width half maximum), while the lower one is at
the resolution of the Planck satellite, about $5$ arcminutes in the
best channel. The units are microKelvins. Note the scale change
between the upper and lower figures.  Planck will have nearly 10,000
times as much data as COBE, and because smaller scale fluctuations are
naturally damped, should capture almost all the primary anisotropy
structure that there is to see. The power spectrum encoding the
wavelength structure in this figure is shown in Fig.~\ref{fig:clproj}.}
\label{fig:allsky}
 \end{center}
\end{figure}

\begin{figure}[htbp]
\begin{center}
\plotonecfpa{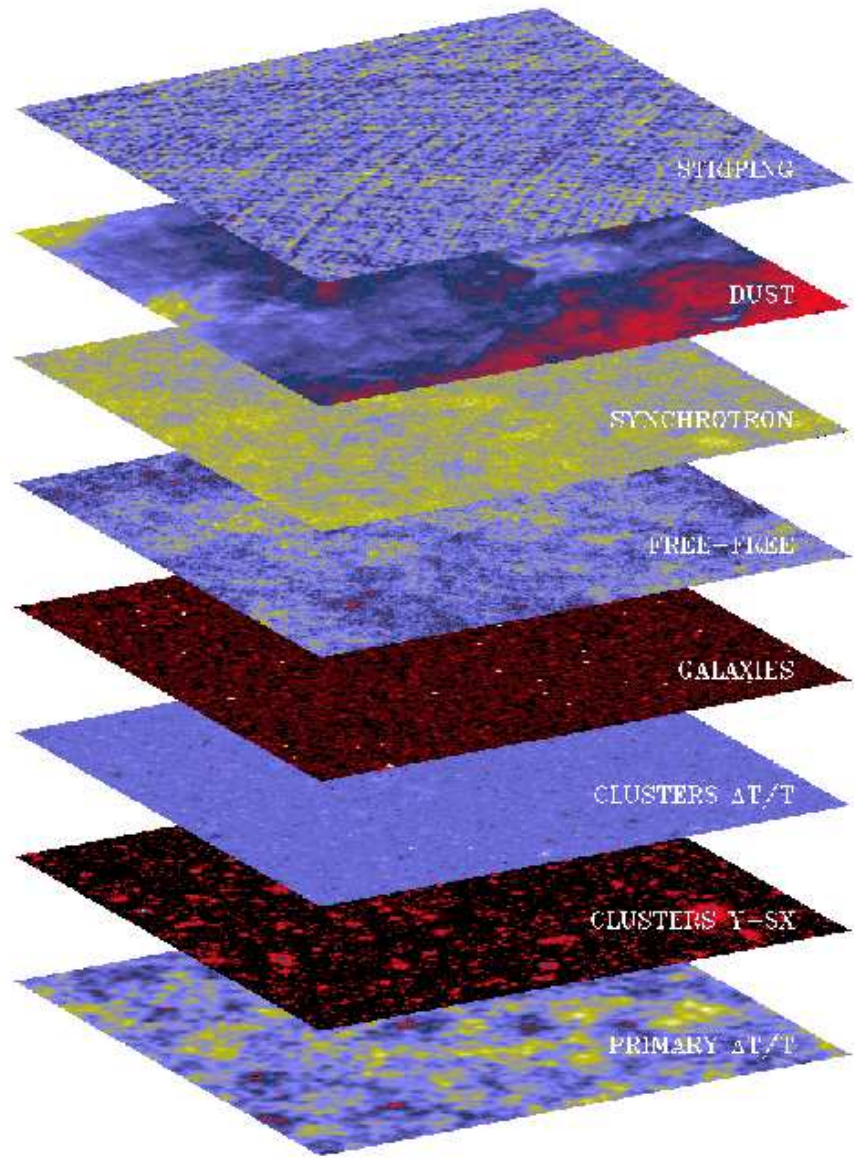}
\caption[Sandwich]{\baselineskip=10pt\small A schematic view of many
of the possible microwave foregrounds that need to be separated from
the primary CMB anisotropy pattern shown in the $10^\circ$ map at the
bottom.  These include noise effects (striping), galactic foregrounds
(dust, synchrotron and bremsstrahlung or free-free emission) as well
as extragalactic foregrounds (radio and infrared galaxies, scattering
by hot gas in clusters, ``Y-SX'' and the Doppler effect arising from
moving clusters, ``$\Delta T/T$'').  Each of these has a unique
temperature pattern on the sky. Sample power spectra are shown in
Fig.~\ref{fig:CLtheory}.}
\label{fig:spatialforegrounds}
\end{center}
\end{figure}

Figure~\ref{fig:allsky} was provided by Kris Gorski and both
computation and visualization have been handled using the
http://www.tac.dk/~healpix software
package. Figure~\ref{fig:spatialforegrounds} was provided by Francois
Bouchet and Richard Gispert. We also thank Julian Borrill and David
Spergel for discussion of computer timings and algorithmic issues.

\end{document}